\documentclass[11pt]{article}
\usepackage[pctex32]{graphics}

\def\ben{\begin{equation}}
\def\een{\end{equation}}

\def\nn{\nonumber} \def\bd{\begin{document}} \def\ed{\end{document}}
\def\ds{\documentstyle} \let\fr=\frac \let\bl=\bigl \let\br=\bigr
\let\Br=\Bigr \let\Bl=\Bigl
\let\bm=\bibitem
\let\na=\nabla
\let\pa=\partial \let\ov=\overline
\newcommand{\be}{\begin{equation}}
\newcommand{\ee}{\end{equation}}
\def\ba{\begin{array}}
\def\ea{\end{array}}
\def\ft#1#2{{\textstyle{\frac{\scriptstyle #1}{\scriptstyle #2} } }}
\def\fft#1#2{{\frac{#1}{#2}}}
\def\del{\partial}
\def\vp{\varphi}
\def\sst#1{{\scriptscriptstyle #1}}
\def\oneone{\rlap 1\mkern4mu{\rm l}}
\def\td{\tilde}
\def\wtd{\widetilde}
\def\ie{{\it i.e.\ }}
\def\dalemb#1#2{{\vbox{\hrule height .#2pt
        \hbox{\vrule width.#2pt height#1pt \kern#1pt
                \vrule width.#2pt}
        \hrule height.#2pt}}}
\def\square{\mathord{\dalemb{6.8}{7}\hbox{\hskip1pt}}}
\newcommand{\ho}[1]{$\, ^{#1}$}
\newcommand{\hoch}[1]{$\, ^{#1}$}
\newcommand{\bea}{\setlength\arraycolsep{2pt} \begin{eqnarray}}
\newcommand{\eea}{\end{eqnarray}}
\newcommand{\ra}{\rightarrow}
\newcommand{\lra}{\longrightarrow}
\newcommand{\Lra}{\Leftrightarrow}
\newcommand{\bp}{\tilde \beta^\prime}
\newcommand{\tr}{{\rm tr} }
\newcommand{\Tr}{{\rm Tr} }
\def\0{{\sst{(0)}}}
\def\1{{\sst{(1)}}}
\def\2{{\sst{(2)}}}
\def\3{{\sst{(3)}}}
\def\4{{\sst{(4)}}}
\def\5{{\sst{(5)}}}
\def\6{{\sst{(6)}}}
\def\7{{\sst{(7)}}}
\def\8{{\sst{(8)}}}
\def\m{{\sst{(m)}}}
\def\n{{\sst{(n)}}}
\def\cA{{{\cal A}}}
\def\cB{{{\cal B}}}
\def\cF{{{\cal F}}}
\def\cG{{{\cal G}}}
\def\cH{{{\cal H}}}
\def\tV{\widetilde V}
\def\tW{\widetilde W}
\def\tH{\widetilde H}
\def\tE{\widetilde E}
\def\tF{\widetilde F}
\def\tA{\widetilde A}
\def\im{{{\rm i}}}
\def\tY{{{\wtd Y}}}
\def\ep{{\epsilon}}
\def\vep{{\varepsilon}}
\def\bD{{{\bar D}}}
\def\R{{{\mathbb R}}}
\def\C{{{\mathbb C}}}
\def\H{{{\mathbb H}}}
\def\CP{{{\mathbb C}{\mathbb P}}}
\def\RP{{{\mathbb R}{\mathbb P}}}
\def\Z{{{\mathbb Z}}}
\def\bA{{{\mathbb A}}}
\def\bB{{{\mathbb B}}}
\def\bC{{{\mathbb C}}}
\def\bD{{{\mathbb D}}}
\def\bE{{{\mathbb E}}}
\def\bZ{{{\mathbb Z}}}
\def\Re{{{\frak{Re}}}}
\def\Im{{{\frak{Im}}}}
\def\cosec{{\,\hbox{cosec}\,}}
\def\Gm{{\Gamma_{\!\! -}}}
\def\Gp{{\Gamma_{\!\! +}}}
\def\stan{{standard }}
\def\nonstan{{supernumerary }}
\def\p{{\partial}}
\def\kdel#1{{\fft{\del}{\del#1}}}

\def\bog{{Bogomolny }}
\def\om{{\hat{\omega}}}

\newcommand{\nnr}{\nonumber \\}
\newcommand{\pd}{\partial}
\newcommand{\ud}{\textrm{d}}
\newcommand{\dTH}{T^{\prime \, 0}_\textrm{H}}
\newcommand{\dOi}{\Omega^{\prime \, 0}_i}
\newcommand{\bx}{{\bf x}}

\begin{document}

\vspace{5mm}
\begin{center}
{\Large \bf Propagations of massive graviton in the deformed
Ho\v{r}ava-Lifshitz gravity } \vspace{12mm}

 \centerline{\large Yun Soo Myung$^{a}$}

\vspace{10mm} {\em Institute of Basic Science and School of
Computer Aided Science \\ Inje University, Gimhae 621-749, Korea}
\vskip .6cm
\end{center}

\begin{center}

\underline{Abstract}
\end{center}

  We   study  massive graviton propagations of  scalar, vector, and tensor modes in the deformed Ho\v{r}ava-Lifshitz gravity
   by introducing Lorentz-violating mass term. It turns out that
   vector and tensor modes
   are massively   propagating on the Minkowski
   spacetime background. However, adding the mass term  does not cure  a ghost instability in the  Ho\v{r}ava scalar.

\vspace{15pt} \baselineskip=18pt
 \noindent $^a$ysmyung@inje.ac.kr

\thispagestyle{empty}

\newpage
\section{Introduction}
Recently Ho\v{r}ava has proposed a renormalizable theory of gravity
at a Lifshitz point~\cite{ho1},  which  may be regarded as a UV
complete candidate for general relativity.  Very recently, the
Ho\v{r}ava-Lifshitz gravity theory has been intensively investigated
in~\cite{ho2,Vis,ho3,VW,klu,Nik,Nas,Iza,Vol,SVW1,CH,CHZ,Nis,KS,OR,Kon,CNPS,SVW,KLM,Cal1,Sak,BPS,BS},
 its cosmological applications in
~\cite{cos1,cos2}, and its black hole solutions in
~\cite{bh1,bh2}.

There are  four versions of Ho\v{r}ava-Lifshitz gravity in the
literature: with/without the detailed balance condition and
with/without the projectability condition~\cite{muk}. Ho\v{r}ava has
originally proposed the projectability condition with/without the
detailed balance condition.  We  mention that the IR vacuum of this
theory is anti de Sitter (AdS) spacetimes. Hence, it is interesting
to take a limit of the theory, which leads  to  a Minkowski vacuum
in the IR sector. To this end, one may modify the theory by
including ``$\mu^4R$" and then, take the $\Lambda_W \to 0$
limit~\cite{KS}. This deformed  Ho\v{r}ava-Lifshitz gravity does not
alter the UV properties of the theory, but it changes the IR
properties from AdS vacuum to  Minkowski vacuum. Hence, in order to
see propagations of fields on the Minkowski spacetime background, we
consider the deformed Ho\v{r}ava-Lifshitz gravity without the
detailed balance condition.

Concerning the projectability condition, its role should be dealt
with carefully to identify the propagation of Ho\v{r}ava scalar  on
the Minkowski background. Actually, there exists a close relation
between projectability and scalar degrees of freedom. The
projectability condition requires that the perturbation $A$  of the
lapse function $N$ depends only on time, thus $A=A(t)$. This implies
that  the $A$-perturbation is  not a Lagrange multiplier (field) but
a time-dependent parameter. This is the key of the theory.

 An urgent issue of the deformed Ho\v{r}ava-Lifshitz
gravity is to answer to the question of whether it can
 accommodate the Ho\v{r}ava scalar $\psi$,
  in addition to two degrees of freedom (DOF) for a massless graviton.
We would like to mention  two relevant works. The
authors~\cite{CNPS} have shown that without the projectability
condition, the Ho\v{r}ava scalar $\psi$ is related to a scalar
degree of freedom appeared in the massless limit of a massive
graviton. This is reminiscent of Fierz-Pauli massive
gravity~\cite{Aub} in which the longitudinal scalar becomes strongly
coupled as $m\to 0$, leading to the vDVZ discontinuity~\cite{vDVZ}.
They argued that perturbative general relativity  cannot be
reproduced in the IR-limit of deformed Ho\v{r}ava-Lifshitz gravity
because of the strong coupling problem. With the projectability
condition, on the other hand, the authors~\cite{SVW} have argued
that $\psi$ is propagating around the Minkowski space but it has a
negative kinetic term, showing a ghost mode.  Moreover, it was found
that
 the Ho\v{r}ava scalar is a ghost if the sound speed squared is
 positive~\cite{BPS}.

In order to understand better the problems arising when one
 modified the gravity in the Lorentz-invariant way, it
was instructive to consider first the Lorentz-invariant massive
gravity by adding  the Fierz-Pauli mass term. However, this term is
not suitable for studying scalar propagations under the
projectability condition.  We remind the reader that the deformed
Ho\v{r}ava-Lifshitz gravity is a Lorentz-violating gravity. Hence
the Lorentz-violating mass terms are more attractive to study the
issue on the propagation of Ho\v{r}ava scalar in the deformed
Ho\v{r}ava-Lifshitz gravity.

 In this work, we investigate  massive graviton propagations of  scalar, vector, and tensor modes
 in the deformed Ho\v{r}ava-Lifshitz gravity under the
 projectability condition
  by introducing Lorentz-violating mass terms.

\section{Deformed Ho\v{r}ava-Lifshitz gravity}
First of all, we introduce the ADM formalism where the metric is
parameterized as
\be ds_{ADM}^2= - N^2  dt^2 + g_{ij} \Big(dx^i - N^i dt\Big)
\Big(dx^j - N^j dt\Big)\,, \ee
Then, the Einstein-Hilbert action can be expressed as
\be \label{Eins} S^{EH} = \fft{1}{16\pi G} \int d^4x \sqrt{g} N
\Big(K_{ij} K^{ij} - K^2 + R \Big)\,, \ee
where $G$ is Newton's constant and extrinsic curvature $K_{ij}$
takes the form
\be K_{ij} = \fft{1}{2N} \Big(\dot g_{ij} - \nabla_i N_j -
\nabla_j N_i\Big)\,. \ee
Here, a dot denotes a derivative with respect to $t$(
$``~\dot{}~"=\frac{\partial}{\partial t}$).

On the other hand, a deformed action of the Ho\v{r}ava-Lifshitz
gravity  is given by~\cite{KS}
\bea%
\label{act}S^{dHL}&=&\int dtd^3\bx\, \Big({\cal L}_0 +\sqrt{g}N\mu^4R + {\cal L}_1\Big)\,,\\
{\cal L}_0 &=& \sqrt{g}N\left\{\frac{2}{\kappa^2}(K_{ij}K^{ij}
\label{action1}-\lambda K^2)+\frac{\kappa^2\mu^2(\Lambda_W R
  -3\Lambda_W^2)}{8(1-3\lambda)}\right\}\,,\\ {\cal L}_1&=&
\sqrt{g}N\left\{\frac{\kappa^2\mu^2 (1-4\lambda)}{32(1-3\lambda)}R^2
-\frac{\kappa^2}{\eta^4} \left(C_{ij} -\frac{\mu
\eta^2}{2}R_{ij}\right) \left(C^{ij} -\frac{\mu
\eta^2}{2}R^{ij}\right) \right\}\,.\label{action2}
\eea%
Here $C_{ij}$ is the Cotton tensor defined by
\be C^{ij}=\epsilon^{ik\ell}\nabla_k\left(R^j{}_\ell
-\frac14R\delta_\ell^j\right)\label{def.K.C} \ee which is obtained
from the variation of gravitational Chern-Simons term with coupling
$1/\eta^2$.
 The full
equations of motion were derived in \cite{cos1} and \cite{bh1}, but
we do not write  them  due to the length. Taking a limit of
$\Lambda_W \to 0$ in ${\cal L}_0+\sqrt{g}N\mu^4 R$, we obtain  the
 Einstein-Hilbert action with $\lambda$~\cite{KS}
 \be
S^{EH\lambda}\equiv \int dt d^3x \tilde{\cal L}_0=\int dt d^3x
\sqrt{g} N\Bigg[\frac{2}{\kappa^2}\Big(K_{ij}K^{ij}-\lambda
K^2\Big)+\mu^4 R\Bigg]\ . \label{SM2} \ee  Comparing
Eq.(\ref{SM2}) with general relativity (\ref{Eins}), the speed of
light and Newton's constant  are given by
\be c^2=\fft{\kappa^2\mu^4}{2},~~
G=\fft{\kappa^2}{32\pi\,c},~~\lambda=1.\label{cg} \ee
Since we consider the $z=3$ Ho\v{r}ava-Lifshitz gravity, scaling
dimensions are  $[t]=-3,[x]=-1, [\kappa]=0,$ $[\mu]=1$, and $[c]=2$.
Even though the scaling dimensions are relevant to the UV
properties, these are also necessary to define the linearized theory
of $z=3$ Ho\v{r}ava-Lifshitz gravity consistently. The reason is
that we have to keep the same dimensions six for all terms, although
couplings of the kinetic term ($2/\kappa^2$) and the sixth order
derivatives ($\kappa^2/2\eta^4$) are dimensionless. To see the UV
property of power-counting renormalizability, it is better to switch
from the $c=1$ units to
 (\ref{cg}) units that impose the scaling dimensions. Switching back to $c=1$
units leads to the case that  is  more suitable for discussing the
IR properties of strong coupling problem and vDVZ discontinuity.

The deformed Lagrangian which is relevant to our study takes the
form~\cite{KS} \bea
\label{delag1} \tilde{{\cal L}}\equiv\tilde{\cal L}_0&+&{\cal L}_1  \\
\label{delag2}=\sqrt{g}N
\Bigg[\frac{2}{\kappa^2}\Big(K_{ij}K_{ij}-\lambda
K^2\Big)&+&\mu^4\Big(R
+\frac{1}{2\omega}\frac{4\lambda-1}{3\lambda-1}R^2-\frac{2}{\omega}R_{ij}R_{ij}\Big) \\
\label{delag3}&+&\frac{\kappa^2\mu}{2\eta^2}\epsilon^{ijk}R_{il}\nabla_jR^l_k
-\frac{\kappa^2}{2\eta^4}C_{ij}C_{ij}\Bigg] \eea where a
characterized parameter $\omega$ is given by \be \omega=\frac{16
\mu^2}{\kappa^2}=\frac{16\sqrt{2}c}{\kappa^3}. \ee   Actually, the
Lagrangian (\ref{delag2}) is enough to describe  scalar and vector
propagations because (\ref{delag3}) from the Cotton tensor
contributes to the tensor propagations only. For $\lambda=1$, taking
the IR-limit of $\omega \to \infty$ while keeping  $c^2=1$ fixed  is
equivalent to recovering the Einstein gravity. Explicitly, this
limit implies $\kappa^2 \to 0(\mu^4\sim \kappa^{-2} \to \infty)$
which means that the kinetic term and curvature term $\mu^4R$
dominate over all higher order curvature terms. The deformed
Lagrangian (\ref{delag1}) can be redefined to be \be \label{lkp}
\tilde{\cal L}={\cal L}_K+{\cal L}_V, \ee where ${\cal L}_K({\cal
L}_V)$ denote the kinetic (potential) Lagrangian.

 We wish to consider
perturbations of the metric around Minkowski spacetimes, which is
a solution to the full theory (\ref{delag1})
 \be \label{decom1}
g_{ij}= \delta_{ij}+\eta h_{ij},~N= 1+ \eta n,~N_i= \eta n_i, \ee
where a dimensionless coupling constant $\eta$ from gravitational
Chern-Simons term is included to define the perturbation. The
inclusion of $\eta$ makes sense because the noninteracting limit
corresponds to sending $\eta \to 0$ while keeping the ratio
$\gamma=\kappa/\eta$ fixed~\cite{ho1}. This in turn provides the
IR-limit of $\kappa \to 0(\omega \to \infty)$.  For $\lambda=1$,
this limit yields a one-parameter family of free-field fixed points
parameterized by $\gamma$.

At quadratic order the action (\ref{SM2}) turns out to be \bea
\label{EHlambda} S^{EH\lambda}_2 &=& \eta^2\int dt d^3x \Bigg\{{1
\over \kappa^2} \left[{1\over 2} \dot h_{ij}^2 -{\lambda\over 2}
\dot h^2 + (\partial_i n_j)^2 +(1-2\lambda) (\partial \cdot n)^2 - 2
\partial_i n_j(\dot h_{ij} -\lambda \dot h \delta_{ij})\right]
\nonumber\\
&&\phantom{x} +  {\mu^4\over 2} \left[ -\frac{1}{2}(\partial_k
h_{ij})^2+\frac{1}{2}(\partial_i h)^2 +(\partial_i
h_{ij})^2-\partial_i h_{ij}\partial_j h + 2 n (\partial_i
\partial_j h_{ij}-\partial^2 h) \right]\Bigg\}  \eea
with $h=h_{ii}$.  A general Lorentz-violating  (LV) mass term is
given by~\cite{Rub,RT} \be
\label{mass1}S_2^{LV}=\frac{\eta^2}{2\kappa^2} \int dt d^3x \Bigg\{
4m_0^2 ~n^2+2m_1^2~n_i^2-\tilde{m}_2^2~h_{ij}^2+\tilde{m}_3^2~
h^2+4\tilde{m}_4^2~ nh\Big\}. \ee As was pointed out  in \cite{dub},
it provides  various phases of mass gravity in general relativity.
In this work, we add  Eq.(\ref{mass1}) to the linearized theory of
deformed Ho\v{r}ava-Lifshitz gravity to investigate the instability
of Ho\v{a}ava scalar and  strong coupling problem. In this work, we
choose the  case of $m_0=0$ and $\tilde{m}_4=0$ because  the lapse
parameter $n(t)$ enters these terms. At this stage, we would like to
mention that for generic backgrounds, the case of $m_1=0$ has
provided  a well-defined case in bigravity and massive
gravity~\cite{RT,BCNP}. Also, the generic case could be well behaved
in generic backgrounds~\cite{BCNP}. We compare (\ref{mass1})  with
the Lorentz-invariant Fierz-Pauli mass term as~\cite{FP}
 \be \label{mass2}
S_2^{FP}=\frac{ \eta^2}{2\kappa^2} \int dt  d^3x \Bigg\{- m^2
h_{\mu\nu}h^{\mu\nu}+ m^2 \Big(h^\mu~_\mu \Big)^2\Bigg\}. \ee

 In order to analyze  physical propagations thoroughly, it is convenient to use the cosmological
decomposition in terms of scalar, vector, and tensor modes under
spatial rotations $SO(3)$~\cite{MFB}
 \bea \label{pert}
 n &=&-\frac{1}{2}A,\nn \\
 n_i&=&\Big(\partial_iB+V_i\Big),\label{decom2} \\
 h_{ij}&=&\Big(\psi\delta_{ij}+\partial_i\partial_j E+2\partial_{(i}F_{j)}+t_{ij}\Big), \nn \eea
where the conditions of
$\partial^iF_i=\partial^iV_i=\partial^it_{ij}=t_{ii}=0$ are
imposed.
 The
last two conditions mean that $t_{ij}$ is a transverse and
traceless tensor in three spatial dimensions.  Using this
decomposition, the scalar modes ($A,B,\psi,E$), the vector modes
($V_i,F_i$), and the tensor modes ($t_{ij}$) decouple completely
from each other. These all amount to 10 degrees of freedom for a
symmetric tensor in four dimensions.

Before proceeding, let us check dimensions. This is a necessary step
to obtain a consistently massive linearized theory. We observe that
$[n]=0,~[n_i]=2,$ and $[h_{ij}]=0$, which imply
$[A]=0,~[B]=1,~[V_i]=2,~[\psi]=0,~[E]=-2,~[F_i]=-1,$ and
$[t_{ij}]=0$.  Also, the masses take  scaling dimensions:
$[m_1^2]=2$ and $[\tilde{m}_2^2]=[\tilde{m}_3^2]=[\tilde{m}_4^2]=6$.
In order to find the true mass with dimension 1, we redefine mass
squares as \be \tilde{m}_i^2=c^2 m^2_i, ~~{\rm for }~~ i=2,3,4 \ee
which implies that \be [m_2^2]=[m_3^2]=[m_4^2]=2.\ee The Fierz-Pauli
mass term is recovered when all masses  are equal except for $m_0$
as \be \label{mass3} m_1^2=m_2^2=m_3^2=m_4^2=m^2;~m_0=0. \ee Hence,
the Fierz-Pauli mass term is not suitable for studying massive
scalar propagations with projectability condition because the latter
condition implies that $n(t)$ is not a field  and thus, it requires
$ m_4^2=0$.

 The
bilinear action is obtained by substituting (\ref{decom2}) into the
quadratic action (\ref{EHlambda}) as
 \bea \label{fehl}
  S^{EH\lambda}_2 =\frac{1}{2\gamma^2}
 \int dtd^3x &\Bigg\{&
  \Big[  3(1-3\lambda)\dot{\psi}^2
+2\partial_iw_j\partial^iw^j
-4\Big((1-3\lambda)\dot{\psi}+(1-\lambda)\partial^2\dot{E}\Big)\partial^2B
\nonumber\\
&&+4(1-\lambda)(\partial^2B)^2+2(1-3\lambda)\dot{\psi}\partial^2\dot{E}
+(1-\lambda)(\partial^2\dot{E})^2+ \dot{t}_{ij}\dot{t^{ij}}\Big]
\nonumber\\
&& +c^2\Big(2\partial_k\psi\partial^k\psi +4A\partial^2\psi
-\partial_k t_{ij}\partial^k t^{ij}\Big)\Bigg\} \eea with
$\gamma^2=\kappa^2/\eta^2$ and $w_i=V_i-\dot{F}_i$. We would like to
point out the coupling of $\frac{1}{2\gamma^2}$ in the front of the
quadratic action because we have chosen the perturbations
(\ref{decom1}).
 The
higher order action obtained from ${\cal L}_1$ takes the form \be
S^1_2=\frac{\kappa^2\mu^2\eta^2}{8}\int dt d^3x \Bigg\{
-\frac{1-\lambda}{2(1-3\lambda)} \psi
\partial^4 \psi -\frac{1}{4}t_{ij}\partial^4 t^{ij}
+\frac{1}{\mu \eta^2} \epsilon^{ijk} t_{il} \partial^4
\partial_j t^l~_k+\frac{1}{\mu^2\eta^4}
t_{ij} \partial^6 t^{ij}
 \Bigg\}. \label{1action} \ee
We find  that  two modes of scalar $\psi$ and tensor $t_{ij}$ exist
in the higher order action only, missing vector modes. Since the
spatial slice is  conformally flat, the vanishing Cotton tensor and
the absence of six derivative term result in the scalar sector.
Also, the Cotton tensor does not contribute to vector modes
($V_i,\dot{F}_i$). The vectors are decoupled completely from
bilinear terms of the potential ${\cal L}_V$. This is because the
vector belongs to gauge degrees of freedom in the massless graviton
theory, while it has 2 DOF in the massive graviton theory. Hence,
the disappearance of vector is natural  for the massless theory of
$z=3$ Ho\v{r}ava-Lifshitz gravity.

 Now we are in a
position to discuss the diffeomorphism in the $z=3$
Ho\v{r}ava-Lifshitz gravity. Since the anisotropic scaling of
temporal and spatial coordinates ($t\to b^z t, x^i \to b x^i$), the
time coordinate $t$ plays a privileged role. Hence, the spacetime
symmetry is smaller than the full diffeomorphism (Diff) in the
general relativity~\cite{ABGV}. The quadratic action of
$S^{EH\lambda}_2+S^1_2$ should  be  invariant under the
``foliation-preserving" diffeomorphism (FDiff) whose transformation
is given by \be t \to \tilde{t}=t+\epsilon^0(t),~~x^i \to
\tilde{x}^i=x^i+\epsilon^i(t,\bf{x}). \ee Using the notation of
$\epsilon^\mu=(\epsilon^0,\epsilon^i)$ and
$\epsilon_\nu=\eta_{\nu\mu}\epsilon^\mu$, the perturbation of metric
transforms as \be \delta g_{\mu\nu} \to
\delta\tilde{g}_{\mu\nu}=\delta g_{\mu\nu}+\partial_\mu
\epsilon_\nu+\partial_\nu \epsilon_\mu. \ee Further, making a
decomposition $\epsilon^i$ into a scalar $\xi$ and a pure vector
$\zeta^i$ as $\epsilon^i=\partial^i\xi+ \zeta^i$ with $\partial_i
\zeta^i=0$, one finds the transformation for scalars \be
\label{trans1} A(t) \to \tilde{A}(t)=A(t)-2\dot{\epsilon^0}(t),~\psi
\to \tilde{\psi}=\psi,~ B \to \tilde{B}=B+\dot{\xi},~E \to
\tilde{E}=E+2\xi.\ee On the other hand, the vector and the tensor
take the forms \be \label{trans2} V_i \to \tilde{V}_i=V_i
+\dot{\zeta}_i,~F_i\to \tilde{F}_i=F_i+\zeta_i,~t_{ij} \to
\tilde{t}_{ij}=t_{ij}. \ee Considering scaling dimensions of
$[\epsilon^0]=-3$ and $[\epsilon^i]=-1$,  we have $[\xi]=-2$ and
$[\zeta^i]=-1$. For the FDiff transformations,  gauge invariant
combinations are \be t_{ij},~~w_i=V_i-\dot{F}_i, \ee for tensor and
vector modes, respectively and \be \psi,~~\Pi=2B-\dot{E} \ee for two
scalar modes. At this stage, we note scaling dimensions: $[w_i]=2$
and $[\Pi]=1$.

Let us  express the quadratic action (\ref{fehl}) in terms of
gauge-invariant quantities as~\cite{KLM} \bea S^{EH\lambda}_2
=\frac{1}{2\gamma^2}
 \int dtd^3x &\Bigg\{&
\Big[  3(1-3\lambda)\dot{\psi}^2 -2w_i\bigtriangleup w^i
-2(1-3\lambda)\dot{\psi}\bigtriangleup \Pi+(1-\lambda)(\bigtriangleup\Pi)^2 \nonumber \\
\label{secondlam}&+& \dot{t}_{ij}\dot{t}^{ij}\Big]
+c^2\left(-2\psi\bigtriangleup\psi +4A(t)\bigtriangleup \psi +
t_{ij} \bigtriangleup t^{ij}\right)\Bigg\} \eea with the spatial
Laplacian $\bigtriangleup=\partial^2$. We note that $S_2^1$ in
(\ref{1action}) contains only $\psi$ and $t_{ij}$, which are
 gauge-invariant quantities. We emphasize that ``$A(t)$" leaves a
gauge-dependent quantity alone. Thus,  it seems that if
$A(t)\not=0$, one does not  obtain the gauge-invariant quadratic
action $S_2^{EH\lambda}$. However, ``$4A(t)\bigtriangleup \psi$" is
a surface term and thus,  we drop it from studying the propagations
of massive graviton.

Finally, the LV mass term (\ref{mass1}) leads to
 \bea
\label{mass4}S_2^{LV}=\frac{1}{2\gamma^2} \int dt  d^3x \Bigg[2
m_1^2\Big(V_i^2+(\partial_iB)^2\Big)
&-&\tilde{m}_2^2\Big(t_{ij}t^{ij}+2(\partial_iF_j)^2+(\partial_i\partial_j E)^2+2 \psi \partial^2 E+3 \psi^2\Big) \nn \\
&+&\tilde{m}_3^2\Big( \partial^2 E+3\psi\Big)^2 \Bigg]. \eea which
is not invariant under FDiff transformations because we could not
express whole terms in terms of gauge-invariant quantities.

\section{Massive tensor and vector propagations}
Before proceeding, we  conjecture that out of the 5 DOF of the
massive graviton, 2 of these are expressed as transverse and
traceless tensor modes $t_{ij}$, 2 of these are expressed as
transverse vector modes $F_i$, and the remaining one is from
Ho\v{r}ava scalar  $\psi$.
\subsection{Tensor modes} The field equation for tensors is given
by \be \label{tensormo} \ddot{t}_{ij}-c^2 \bigtriangleup t_{ij}
+c^2m^2_2 t_{ij}+\frac{2c^2}{\omega}\bigtriangleup^2t_{ij}-
\frac{\kappa^4\mu}{4\eta^2}\epsilon_{ilm}\partial^l\bigtriangleup^2t_j~^m-
\frac{\kappa^4}{4\eta^4} \bigtriangleup^3 t_{ij}=0. \ee The
requirement that these modes are not tachyonic gives the stability
condition \be m^2_2\ge 0. \ee In the absence of mass, these modes
describe the chiral primordial gravitational waves~\cite{BS,Myung4}.
These circularly polarized modes are possible because the Cotton
tensor $C_{ij}$ is present, making parity violation. In the presence
of a mass term, it may describe massive chiral gravitational waves.

 \subsection{Vector modes}

 It is clear
from Eqs.(\ref{fehl}) and (\ref{mass4}) that $V_i$ enters the action
without temporal derivatives, that is, it is a non-dynamical field
in the massless theory.  A massive vector Lagrangian takes the form
\be \label{vector1} {\cal L}^v=\frac{1}{\gamma^2} \Bigg[
-w_i\bigtriangleup w^i+m^2_1 V_i^2-\tilde{m}_2^2 (\partial_iF_j)^2
\Bigg] \ee with $w_i=V_i-\dot{F}_i$. It is clear that in the absence
of mass terms, $w_i$ is a nonpropagating vector mode. We integrate
$V_i$ out using the field equation obtained by varying action with
respect to $V_i$  \be \bigtriangleup (V_i-\dot{F}_i)-m_1^2 V_i=0 \ee
which implies \be V_i=\frac{\bigtriangleup}{\bigtriangleup-m_1^2}
\dot{F}_i. \ee Plugging this expression into Eq.(\ref{vector1})
leads to be \be \label{vector4} {\cal L}^v=\frac{1}{\gamma^2} \Bigg[
\frac{\bigtriangleup m_1^2}{\bigtriangleup -m_1^2} \dot{F}_i^2+
\tilde{m}_2^2 F_i \bigtriangleup F^i \Bigg]. \ee Considering $
\bigtriangleup<0$, the time kinetic term is always positive. In this
case,  we introduce a canonical vector field $\tilde{F}_i$ to obtain
a canonical action  as  \be \label{cvf}
F_i=\frac{\gamma}{m_1}\sqrt{\frac{\bigtriangleup
-m_1^2}{2\bigtriangleup }} \tilde{F}_i \propto \frac{1}{m_1
M_{Pl}}\sqrt{\frac{\bigtriangleup -m_1^2}{2\bigtriangleup }}
\tilde{F}_i\ee in the  $c=1$  units.  Then, the Lagrangian
(\ref{vector4}) takes the canonical form  \be {\cal
L}_c^v=\frac{1}{2} \Bigg[ \dot{\tilde{F}}_i^2-\frac{m_2^2}{m_1^2}
(\partial_i\tilde{F}_j)^2-m_2^2 \tilde{F}_j^2 \Bigg]. \ee  Now let
us discuss the strong coupling issue. In order to discuss the strong
coupling problem, we first note that \be \frac{1}{8\pi
G}=\frac{4c}{\kappa^2}\equiv M^2_{Pl}, \ee which leads to an
important relation between $\gamma$ and Planck mass scale $M_{Pl}$
\be \label{imrel} \gamma=\frac{2\sqrt{c}}{\eta M_{Pl}} \propto
\frac{1}{M_{Pl}} \ee in the $c=1$ units.  Considering the relation
Eq.(\ref{cvf}), the original vector field is proportional to
$(mM_{Pl})^{-1}$ and from Eq.(\ref{vector1}), the gauge-invariant
combination $w_i$ takes the form \be w_i \propto \frac{m}{M_{Pl}}
\tilde{F}_i \ee which show that vector modes at small $m$ is
precisely the same as in the Fierz-Pauli case.  The analysis in
Ref.\cite{AGS} suggests the strong coupling occurs at $E \sim
\sqrt{mM_{Pl}}$, which is a  high scale. Its equation of motion is
given by \be \ddot{\tilde{F}}_{i}-\frac{m_2^2}{m_1^2} \bigtriangleup
\tilde{F}_{j} +m^2_2 \tilde{F}_{i}=0. \ee The above leads to the
dispersion relation  \be p_0^2=\frac{m^2_2}{m_1^2}p^2+m^2_2, \ee
where \be \frac{\partial}{\partial(ct)}\equiv~ ^\prime \to
-p_0,~~\frac{\partial}{\partial x^i} \to -p_i. \ee  For $m_1^2>0$
and $m_2^2>0$, it is obvious that there is no ghosts.

In the Fierz-Pauli case of $m_1^2=m_2^2$, the massive vector
equation reduces to \be \Big(\square -m^2 \Big) \tilde{F}_{i}=0 \ee
which represents a massive vector with two degrees of freedom. Here
$\square=-\partial_0^2+\bigtriangleup$.

\section{Massive scalar propagations}
It turned out that for $1/3<\lambda<1$, there is ghost instability
for the Ho\v{r}ava scalar~\cite{BS}. Thus, our primary concern is to
investigate whether adding a LV mass term can cure this instability.
The scalar Lagrangian composed of $\psi,~B,$ and $E$ takes the form
\bea \label{scalarL} {\cal L}^s =\frac{1}{2\gamma^2} &\Bigg[&-
3(3\lambda-1)\dot{\psi}^2 +2(3\lambda-1)\dot{\psi}\bigtriangleup
(2B-\dot{E})-(\lambda-1)\Big(\bigtriangleup(2B-\dot{E})\Big)^2\nn
\\
&-&2c^2\psi \bigtriangleup
\psi-\frac{(1-\lambda)}{2(3\lambda-1)}\frac{4c^2}{\omega}\psi
\bigtriangleup^2\psi -2m_1^2 B\bigtriangleup B
-\tilde{m}_2^2\Big(E\bigtriangleup^2E +2 \psi \bigtriangleup E +3
\psi^2\Big) \nn
\\
&+&\tilde{m}^2_3\Big(\bigtriangleup E+ 3\psi\Big)^2\Bigg]. \eea
Variations with respect to $B$ and $E$ lead to \bea \label{scalarB}
&&(3\lambda-1)\dot{\psi}+(\lambda-1)\bigtriangleup
\dot{E}-2(\lambda-1)\bigtriangleup B-m^2_1B=0, \\
\label{scalarE}&&(3\lambda-1)\ddot{\psi}-(\lambda-1)\bigtriangleup(2\dot{B}-\ddot{E})+(\tilde{m}^2_3-\tilde{m}_2^2)\bigtriangleup
E+( 3\tilde{m}^2_3-\tilde{m}_2^2)\psi=0\eea which show complicated
relations between three fields. In this case, the diagonalization
process seems to a formidable task.  Hence, we consider three cases
of massless, $B=0$ and $E=0$.

\subsection{Massless case and Strong coupling problem}  In the massless case,  Eqs.(\ref{scalarB}) and (\ref{scalarE}) reduces to
a single relation \be \bigtriangleup\Pi=\frac{1-3\lambda}{1-\lambda}
\dot{\psi} \ee with $\Pi=2B-\dot{E}$. Substituting this into
Eq.(\ref{scalarL}) with $m_1^2=m_2^2=m_3^2=0$, one finds the
Ho\v{r}ava Lagrangian for $\psi$ \be  \label{scalareL}{\cal
L}^s_{H}=\frac{1}{2\gamma^2}\Bigg[\frac{2(1-3\lambda)}{1-\lambda}
\dot{\psi}^2 -2c^2 \psi \bigtriangleup
\psi-\frac{1-\lambda}{2(1-3\lambda)}\frac{4c^2}{\omega} \psi
\bigtriangleup^2 \psi \Bigg]. \ee We note that the case of
$E=0$-gauge leads to Eq.(\ref{scalareL}) exactly. Here, it is
obvious that for $\frac{1}{3} <\lambda <1$, the time kinetic term
becomes negative and thus, the Ho\v{r}ava scalar suffers from the
ghost instability. In addition, comparing it with the tensor
Lagrangian indicates  that the second term is opposite and the third
term is consistent with the tensor term.  The dispersion relation is
given by \be p_0^2=-\frac{1-\lambda}{2(1-3\lambda)}
p^2+\frac{(1-\lambda)^2}{2(1-3\lambda)^2}\frac{4}{\omega} p^4. \ee

Now let us mention the strong coupling problem. Introducing the
sound speed squared $c_\psi^2$ as \be
c_\psi^2=\frac{1-\lambda}{3\lambda-1}, \ee the  Ho\v{r}ava
Lagrangian can be rewritten to be \be \label{hslag}{\cal
L}^s_{H}=-\frac{c^2}{\gamma^2}\Bigg[\frac{1}{c^2_\psi}
(\psi^\prime)^2 + \psi \bigtriangleup \psi-\frac{c^2_\psi}{\omega}
\psi \bigtriangleup^2 \psi \Bigg]\ee which is exactly the same form
as in Ref.\cite{KA} for $\gamma^2=c^2=1$ and ignoring the fourth
order derivative  term. For $1/3<\lambda<1$, $c^2_\psi>0$ but the
time kinetic term is negative definite (ghost).

On the other hand, if the Ho\v{r}ava scalar $\psi$ is not a ghost,
then it is unstable because of $c^2_\psi<0$~\cite{BPS}. considering
the canonical scalar $\tilde{\psi}$ with $c^2=1$ \be \label{res}
\psi=\frac{\gamma |c_\psi|}{\sqrt{2}}\tilde{\psi}, \ee  the above
scalar Lagrangian takes the canonical form \be {\cal L}^c_H=\Bigg[
\frac{1}{2} (\tilde{\psi}^\prime)^2-\frac{c^2_\psi}{2} \tilde{\psi}
 \bigtriangleup \tilde{\psi}+ \frac{c^4_\psi}{2\omega}\tilde{\psi}
 \bigtriangleup^2 \tilde{\psi} \Bigg]. \ee
 In this
case, we need to take into account the last  term of
Eq.(\ref{hslag}) to address the fate of the instability. Here the
time scale of the instability is at least
$\frac{\gamma}{|c_\psi|}\propto \frac{1}{|c_\psi|M_{Pl}}$. In order
not to have the instability within the age of the universe ($1/H_0$
with the present Hubble parameter $H_0$), one needs to have
$|c_\psi| \sim H_0/M_{Pl}$ which means that $|c_\psi|\to 0$ (
$\lambda\to 1$) or the UV scale of the theory is very
low~\cite{BPS}.  It is known that if $c^2_\psi$ is small, the higher
order interactions (for example, cubic interactions of $\psi$)
become increasingly important. For simplicity, let us consider the
scalar sector $n_i=\partial_i B$ and $h_{ij}=e^{\psi}\delta_{ij}$
with $E=0$-gauge. Then, the constraint takes the form \be
\label{newbp} \bigtriangleup B=\frac{1-3\lambda}{2(1-\lambda)}
\dot{\psi}=-\frac{1}{2c_\psi^2} \dot{\psi}=-\frac{c}{2c_\psi^2}
\psi^\prime. \ee The third order Lagrangian is given by~\cite{KA}
\be {\cal L}^s_3 \propto \frac{c^2}{\gamma^2} \Bigg[ \psi \partial_i
\psi
\partial^i \psi -\frac{3}{c^2_\psi} \psi
(\psi^\prime)^2+\frac{3\psi}{2c^2}\Big(\partial_i\partial_j
B\partial^i\partial^j B-(\bigtriangleup
B)^2\Big)-\frac{2}{c^2}\bigtriangleup B
\partial_i \psi \partial^iB \Bigg]. \ee
Plugging Eq.(\ref{newbp}) into ${\cal L}^s_3$ leads to \be {\cal
L}^s_3 \propto \frac{c^2}{\gamma^2} \Bigg[ \psi \partial_i \psi
\partial^i \psi -\frac{3}{c^2_\psi} \psi
(\psi^\prime)^2+\frac{3\psi}{8c^4_\psi}\Big(\frac{\partial_i\partial_j}{\bigtriangleup}
\psi^\prime \frac{\partial^i\partial^j}{\bigtriangleup}
\psi^\prime-(\psi^\prime)^2\Big)-\frac{2}{c^4_\psi}\psi^\prime
\partial_i \psi \frac{\partial^i}{\bigtriangleup}\psi^\prime\Bigg]. \ee
Finally, using (\ref{res}), we have the canonical third order
Lagrangian~\cite{PS} \be {\cal L}^c_3 \propto \frac{1}{2\sqrt{2}}
\Bigg[ \frac{c^3_\psi}{ M_{Pl}} \tilde{\psi}
\partial_i \tilde{\psi}
\partial^i \tilde{\psi} -\frac{3}{ M_{Pl}} \tilde{\psi}
(\tilde{\psi}^\prime)^2+\frac{3\tilde{\psi}}{8c_\psi
 M_{Pl}}\Big(\frac{\partial_i\partial_j}{\bigtriangleup}
\tilde{\psi}^\prime \frac{\partial^i\partial^j}{\bigtriangleup}
\tilde{\psi}^\prime-(\tilde{\psi}^\prime)^2\Big)-\frac{2}{c_\psi
 M_{Pl}}\tilde{\psi}^\prime
\partial_i \tilde{\psi} \frac{\partial^i}{\bigtriangleup}\tilde{\psi}^\prime\Bigg]. \ee
We observe that the last two terms scale as $(c_\psi M_{Pl})^{-1}$
and thus, the Ho\v{r}ava scalar becomes  strong coupled for  $c_\psi
\to 0(\lambda \to 1)$. Importantly, we note that all terms which
blow up in that limit come from the kinetic Lagrangian ${\cal L}_K$
in Eq.(\ref{lkp}). This means that  the potential terms cannot cure
the strong coupling problem.

\subsection{$B=0$-gauge case}

In this case, Eq.(\ref{scalarE}) leads to \be
(\lambda-1)\bigtriangleup\ddot{E}+(\tilde{m}^2_3-\tilde{m}_2^2)\bigtriangleup
E+(3\lambda-1)\ddot{\psi}+( 3\tilde{m}^2_3-\tilde{m}_2^2)\psi=0\ee
which seems to be difficult to express $E$ in terms of $\psi$. For
$m_2^2=m_3^2\equiv m^2$ case, the above relation leads to a rather
simple one \be \label{berel}
\bigtriangleup\ddot{E}=-\frac{1-3\lambda}{1-\lambda}
\ddot{\psi}+\frac{2c^2m^2}{1-\lambda}\psi. \ee However, it is not
easy to derive a relation without derivative from (\ref{berel}).
Hence, we could not express $4 c^2 m^2 \psi \bigtriangleup E$ in
Eq.(\ref{scalarL}) in terms of $\psi$.

\subsection{$E=0$-gauge case}

In this case, we require $m_3^2=0$ for the consistency. For
$m_3^2\not=0$, the mass term of $\psi$ takes
$3c^2(3m_3^2-m_2^2)\psi^2$, which induces a tachyonic mass for $
3m_3^2>m_2^2$. The relation between $B$ and $\psi$ takes the form
\be
B=\frac{(3\lambda-1)\dot{\psi}}{2(\lambda-1)\bigtriangleup+m_1^2}.
\ee The $m_1^2=0$ case leads to the well-known relation of
$B=2\dot{\psi}/m_1$.  Substituting this into Eq.(\ref{scalarL}), we
have the Lagrangian \bea {\cal L}^s_{E=0}=\frac{1}{2\gamma^2}\Bigg[
\dot{\psi}\Bigg(3(1-3\lambda)&+&\frac{4(1-3\lambda)^2\Big[(\lambda-1)\bigtriangleup^2+\frac{m_1^2\bigtriangleup}{2}\Big]}
{[2(1-\lambda)\bigtriangleup+m_1^2]^2}\Bigg)\dot{\psi} \nn \\
&-&
\psi\Bigg(2c^2\bigtriangleup+\frac{1-\lambda}{2(1-3\lambda)}\frac{4c^2\bigtriangleup^2}{\omega}+3c^2m_2^2\Bigg)\psi\Bigg].
\eea It is clear that ${\cal L}^s_{E=0}$ with $m_1^2=m_2^2=0$
recovers ${\cal L}^s_H$ in Eq.(\ref{scalareL}).
\begin{table}
\center
 \caption{Signs for time kinetic terms for massive  Ho\v{r}ava scalar $\psi$ with $\bigtriangleup<0$.}
\vskip .6cm
\begin{tabular}{|c|c|c|}
  \hline
   & $1/3<\lambda<1$ & $\lambda>1$ \\ \hline
  $3(1-3\lambda)$ & $-$ & $-$ \\ \hline
  $(\lambda-1)\bigtriangleup^2$ & $-$ & + \\ \hline
  $\frac{m_1^2\bigtriangleup }{2}$& $-$ & $-$ \\ \hline
  $\frac{2(1-3\lambda)}{1-\lambda}(m_1^2=0)$& $-$ & + \\
  \hline
\end{tabular}
\end{table}
As is shown in Table 1, the mass term contributes to negative term
in the time kinetic terms. Hence, adding the mass term
($m_1^2\bigtriangleup/2$) does not change the ghost instability for
$\frac{1}{3}<\lambda<1$ and further, it may induces the ghost
instability even for $\lambda>1$. The latter case is obviously free
from the ghost instability for the massless case. We show that even
if a Lorentz violating mass term is introduced at the quadratic
level, it could not cure the instability which is present in the
massless case of the Ho\v{r}ava-Lifshitz gravity. The same mass of
$m_1^2=m_2^2$ does not resolve the instability issue.

\section{Discussions}

In order to understand better the problems arising when one
attempts to modify the gravity in the Lorentz-violating way, we
have studied  massive propagations of  scalar, vector, and tensor
modes in the deformed Ho\v{r}ava-Lifshitz gravity
   by introducing Lorentz-violating mass term.
In this approach, we did  choose a
   gauge of $E=0$  to study  massive scalar propagations.
    We have found that tensor
   modes $t_{ij}$ and  vector modes $\tilde{F}_i$ are propagating in
   the Minkowski spacetimes for both mass terms (\ref{mass1}) and (\ref{mass2}).
    However, the propagation of Ho\v{r}ava scalar $\psi$
is not well defined because it still  has  the ghost instability.

We remark that  there exists a strong coupling problem for an
interacting theory of $z=3$ Ho\v{r}ava-Lifshitz gravity beyond the
linearized theory~\cite{SVW,BPS,KA,PS}. In this case, the Ho\v{r}ava
scalar is a ghost if the sound speed squared is positive. In order
to make the scalar graviton healthy, the sound speed squared must be
negative but it is inevitably unstable. Thus, one way to avoid this
is to choose the case that the sound speed squared is  close to
zero, which implies $\lambda \to 1$. However, in the small sound
speed limit, the cubic interactions blows up which means that they
are important at very low energies. This invalidates any linearized
analysis and any predictability is lost due to unsuppressed loop
corrections.

Consequently, the Ho\v{r}ava scalar is still unstable by including a
Lorentz-violating mass term. This implies that the mass terms do not
regularize the bad behavior of the Ho\v{r}ava scalar in the $z=3$
Ho\v{r}ava-Lifshitz gravity which was discussed in~\cite{BPS}.

\section*{Acknowledgement}

This work  was supported by Basic Science Research Program through
the National Research  Foundation (NRF)  of  Korea funded by the
Ministry of Education, Science and Technology (2009-0086861).

\end{document}